\pdfoutput=1
\documentclass{article}

\usepackage[preprint]{nips_2018}

\usepackage{algorithm}
\usepackage{amsmath}
\usepackage[utf8]{inputenc} 
\usepackage[T1]{fontenc}    
\usepackage{hyperref}       
\usepackage{url}            
\usepackage{booktabs}       
\usepackage{amsfonts}       
\usepackage{nicefrac}       
\usepackage{microtype}      
\usepackage[pdftex]{graphicx}
\title{Sampling using Neural Networks for colorizing the grayscale images}

\author{%
  Wonbong Jang\thanks{This paper is based on my dissertation to MSc Computational Statistics and Machine Learning at UCL. Special thanks to Dr. Iasonas Kokkinos for his helpful comments and guidelines. Please reference this \href{http://github.com/wayne1123/colorization}{GitHub} for the code.} \\
  Department of Statistics\\
  London School of Economics\\
  London, WC2A 2AE \\
  \texttt{w.jang@lse.ac.uk} \\
}

\begin{document}

\maketitle
\begin{abstract}
  The main idea of this paper is to explore the possibilities of generating samples from the neural networks, mostly focusing on the colorization of the grey-scale images. I will compare the existing methods for colorization and explore the possibilities of using new generative modeling to the task of colorization. The contributions of this paper are to compare the existing structures with similar generating structures(Decoders) and to apply the novel structures including Conditional VAE(CVAE), Conditional Wasserstein GAN with Gradient Penalty(CWGAN-GP), CWGAN-GP with L1 reconstruction loss, Adversarial Generative Encoders(AGE) and Introspective VAE(IVAE). I trained these models using CIFAR-10 images. To measure the performance, I use Inception Score(IS) which measures how distinctive each image is and how diverse overall samples are as well as human eyes for CIFAR-10 images. It turns out that CVAE with L1 reconstruction loss and IVAE achieve the highest score in IS. CWGAN-GP with L1 tends to learn faster than CWGAN-GP, but IS does not increase from CWGAN-GP. CWGAN-GP tends to generate more diverse images than other models using reconstruction loss. Also, I figured out that the proper regularization plays a vital role in generative modeling.
\end{abstract}
\section{Introduction}

In this section, I will give the brief introduction to this paper, why we are interested in sampling using the neural networks and the generative models, and how it can be applied to the colorization problem. Then, I will elaborate more on how I conducted the experiments.

\subsection{Motivation}

Colorization has been one of the classical problems not only in the field of Computer Vision but also in the area of Art. One of the main difficulty of colorization is to construct the additional information of the images only based on grey-scale images. From this additional generating information, grey-scale images can map into several different colorized images, and this problem is known as multi-modality. For example, a car may be black, green, blue or yellow, but the training set only has one definite answer to this. This multi-modality also has been one of the most critical problems in representative modeling. There are many examples of multi-modality in a real-world application such as colorization or translation.

Neural network has regained its popularity after winning the ImageNet Classification task in 2012. Though its mechanism has not been fully understood, it can classify the several images almost similar or even better than the level of human experts. Recent developments focus on predicting pixels in the image or generating the similar images compared to the original ones. There have been roughly three waves of this trend. One is to sample every pixel using CNN, and the other one is VAE which optimizes its variance lower bound in each step so that the generated images become closer to real images, and the last one is GAN which tries to find the Nash Equilibrium between the Generator network and the Discriminator networks. Now, model developments usually focus on getting the high-quality images as well as stabilizing the training.

Colorization, multi-modal problems, and sampling from generative models have been the active research areas in the last few years. In this trend, I aim to replicate and compare some of the pre-existing models and to try to implement novel hybrid models between VAE and GAN.

\subsection{Aims and Research questions}

In this paper, my aim is to compare the pre-existing generative models using the widely used metric and to explore the possibility of improving the model. In the field of generative modeling, VAE minimizes the element-wise reconstruction error and the regularized error which is the Kullerback Leibler(KL)-divergence between the posterior and the standard Normal prior. It has solid the theoretical background, and we can see the encoded latent spaces directly and control the output images if the latent spaces are sufficiently understood. However, the quality of generated images tends to be blurry. Whereas, GAN is known to produce the sharp images. It minimizes the adversarial loss rather than an element-wise loss. However, its training is not stable, and it is a black-box model. Also, the hybrid models between VAE and GAN are proposed and one trend of the methods is to minimize the reconstruction loss while training GAN, and the other trend is to incorporate the adversarial loss inside VAE by modifying the loss function.

In the field of colorization, previous methods use CNN to colorize the grayscale images and combine VAE and Mixture Density Network(MDN) to generate colourful images from the single grayscale image. Also, there has been the paper using Conditional GAN (CGAN) and reconstruction loss to generate colorized images directly. In this paper, I replicate and modify some of the previous methods in colorization and apply novel approaches in generative modelings for colorization.

One of the problems in comparing the generative modeling is that it does not have an absolute metric to compare the results like the accuracy rate of the test set in the classification task. Widely used quantitative metrics are IS and Frechet Inception Distance(FID). IS is known to be well fit for CIFAR-10 images, and previous works of literature use are to evaluate the performance, so I apply IS to compare the performance of the models.

\subsection{Experiments}
\subsubsection{Data Set}

For the colorization task, any grey-scale images converted from natural images can be the data set. I have used CIFAR-10 and turned them into grey-scale from RGB scale just using the Scikit-Image \textit{rgb2gray} function. CIFAR-10 can be the proper starting point to compare the performance of different architectures because it has been used for quite a while. However, it has a bit small dimension - 32, so for this paper, I enlarge the size of the image to 64. CIFAR-10 has 50,000 training images and 10,000 test images, and it is classified into 10 different categories. 

\subsubsection{Computing Power}

I have used \textit{Amazon AWS p3.2xlarge} which provides Tesla V100 GPU for training the model with the data. I have used \textit{Google Colabarotary} which allows me to use the portion of Tesla K80 GPU for free for developing prototype models. Google Colaboratory has been very useful to train some models with a little data, but it has some limitations to access the memory, and the data is stored for only 12 hours. To train the model, it takes at most 30 hours using \textit{Amazon AWS p3 2x.large}. 

\subsection{Structure of this paper}

In section 2, I will present the theoretical backgrounds and the related concepts in the previous works of neural networks and colorization. These include neural networks, CNN and its application to image classification tasks and its developments, Expectation maximization (EM) and Variational Inferences and sampling methods for approximate inferences, VAE and improvements of VAE, GAN, Wasserstein GAN(WGAN) and their variants. Then, I present some of the existing works of literature on colorization. These include using CNN, CVAE and Mixture Density Networks(MDN), and Conditional GAN. I also present some of the existing quantitative metrics - IS and FID.

From section 3 to section 6, I am going to describe my models and their results. In section 3, I am going to present CNN with L1 and L2 loss. In section 4, I come up with CVAE, and in section 5, I will introduce WGAN-GP and WGAN-GP with L1 reconstruction loss, then in section 6, Adversarial Generative Encoder(AGE), and Introspective VAE(IVAE) would be presented. In those sections, I will explain how I choose the kernel size, the activation function, and explain the new methods to stabilize the training and improve the performance of the models. 

Section 7 is the conclusion of this paper, and I briefly summarize the results and convey critical arguments in this paper, then present some of the remaining future works.

\section{Background and related works}

In this section, I will present some of the ideas about the critical developments in classification tasks and generative modeling. Then, I will review some of the previous papers on colorization applying the different kinds of neural networks, then move onto the quantitative metrics used to evaluate the performances.

\subsection{Neural Network and CNN}

Neural Network is just the composition of linear transforms and non-linear functions, but it is a very powerful function approximator. According to \cite{cybenko1989approximation}, a single hidden layer and a non-linear activation (Sigmoid) function can represent any functions (with enough hidden units). With many more hidden layers, (as we go deeper) the neural network can become more powerful and efficient. (\cite{montufar2014number}).

The learning in a neural network is done by automatic differentiation by backpropagation proposed by \cite{lecun1989backpropagation}. We can compute the loss and get gradients through the forward execution path. Then using the obtained derivatives and errors from forwarding propagation, we can optimize the parameters by gradient descent. Previously, stochastic gradient descent was used widely, but it is quite vulnerable to the saddle points, and the network also suffers from vanishing gradients when it becomes deeper. To cope with these limits, using the momentum such as Nesterov was introduced. Recently, there have been further developments in optimization methods such as RMS Prop by \cite{tieleman2012lecture} and Adam optimizer by \cite{kingma2014adam} to incorporate the idea of using the different orders of momentum better. 

\subsubsection{Convolution Neural Networks and Image Classification}

For image data, \cite{lecun1995convolutional} proposed Convolutional Neural Networks, here convolution layers perform better than Fully-Connected layers. Convolution layer has much smaller number of parameters than fully connected layers, and it enjoys sparse structures. Also, it is similar to how a human's eyes recognize the objects. In human's eyes, we understand the images as the weighted sum of convolutions of images by Gaussian, the first and the second derivative of Gaussian. It is not surprising CNN works well for the visual recognition problems because it follows how human brain system deals with the signal.

\cite{krizhevsky2012imagenet} proposed AlexNet which has five convolution layers and three fully-connected layers. It was the significant breakthrough at ImageNet challenge because the classification error went down from 0.26 to 0.16 in just one year. Before AlexNet, hand-crafted computer vision pipelines won the competition in 2010 and 2011. At AlexNet, the authors use Rectified Linear Unit(ReLU) functions for activation, and apply dropout and weight decay(L2 penalty). Also, they increase the dataset by random augmentation and train six days using 2 GPUs.

\cite{simonyan2014very} proposed VGGNet. It mainly uses 3x3 convolution layer and Fully-Connected layers on the last. It takes approximately 2-3 weeks to train by using 4 GPUs. VGGNet uses ReLU, Dropout and weight decay, and it has 11, 13, 16 and 19 layers.

\cite{szegedy2015going} introduced a family of models(InceptionNet v1 to v4). InceptionNet v1 won the classification task of ImageNet in 2014. The key characteristics of InceptionNet are to use Batch normalization and 1x1 Bottleneck convolution and residual connection and to stack across channels in each Inception module. Batch normalization has stabilized the training process considerably, and 1x1 Bottleneck convolution is very useful to reduce the number of channels. Residual connection works well with vanishing gradients problems. 

\cite{he2016deep} proposed ResNet which won the ImageNet classification contest in 2015. It has the simple design, inspired by VGGNet, but it has become ten times deeper. The problem with the pre-existing models is that the performance doesn't get better even though the model becomes deeper due to vanishing gradients. ResNet solves this problem by using residual connections. It also uses Batch normalization. ResNet-152 achieved the better performance than VGG networks with lower time complexity, and it has the simpler structure than Inception models. 

\cite{yu2015multi} proposed dilated convolutions. The dilated convolutions expand convolution kernel, and we can use only dilated convolutions without using linear and pooling layers. By doing this, it increases the receptive field, and the resolution does not decrease. 

In CNN and an image classification task, there have been so many progress in the last few years. Models become deeper and deeper. New optimizers are introduced like RMSProp and Adam Optimizer, and further activation functions such as ReLU and its variants such as Exponential Linear Unit(ELU) or Leaky Rectified Linear Unit(LeakyReLU) are proposed to deal with gradient vanishing. Batch normalization is very useful to stabilize training and increase the quality of the outputs. Skipping, adding or concatenating the intermediate layers are quite adequate to improve the results. 

\subsection{Generative model - Variational approaches}

The representation learning is to discover the mapping from the representation to output, and also the representation itself, like \textit{autoencoder}. The problem is how to figure out the factors of variations explaining the observed data.

\subsubsection{Expectation Maximization and MDN}

In the unsupervised learning, the probability modeling has been the dominant approach. It focuses on the underlying distributions of the probability distribution. This approach is based on the \textit{Manifold Hypothesis} which states that original data in high dimensional space lies in a low dimensional space. Therefore, if we find the underlying distributions, then we can explain the source of variations in real data.

Before applying the neural network architecture for probability modeling, one of the popular approaches was the Expectation Maximization (EM) which optimizes the lower bound of Free Energy(also known as an evidence lower bound) for all parameters in every step. Free energy of the distribution is supposed to increase every iteration by Jensen's inequality. It is easy to check whether this algorithm works well by just seeing the how free energy rises in every iteration. Therefore, for each iteration, we update the posterior distribution in E-step, then optimize the parameters in M-step, then the free energy keeps increasing before it converges. E-step is the same as minimizing the KL-divergence between the proposed posterior distribution and the original posterior, and M-step is to find the maximum likelihood estimator. However, for EM, we need to know the explicit form of posterior distribution, so we cannot use EM for intractable distributions. 

\cite{bishop1994mixture} proposed Mixed Density Network(MDN) to use a neural network to approximate a function with Mixture of Gaussians. Since the neural network is a powerful function approximator, it can find the posterior distribution of each mixtures and corresponding Gaussians. It can map one input to several corresponding output probabilities, so it may well be fitted for \textit{one to many} cases. However, generating the covariance matrix may suffer the curse of dimension if there are not strict assumptions on the matrix.

\subsubsection{Variational Inference and MCMC}

For intractable distributions, Variational Inference methods such as Variational Bayes has been proposed. Unlike EM, approximate posteriors are not the exact posterior distributions. Increasing free energy does not ensure the convergence to the target distribution, it just finds a close density to the target distribution. Another problem of Variational Inference is that it requires the exact solution even for the simplified cases.

MCMC aims to move towards the more likely regions by repeated sampling and Markov Chains. It guarantees asymptotically exact samples towards target density. However, it is computationally expensive, so it has limitations to scale up to high dimensions. For the large dataset, MCMC requires sampling for every mini-batch, it would be quite slow. 

\begin{equation}
    \mathbb{P}_\theta(z|x) = \frac{\mathbb{P}_\theta(x|z)\mathbb{P}(z)}{\mathbb{P}_\theta(x)}\label{eq:1}
\end{equation}

MCMC and Variational Inference update the posterior given the evidence, as $P_\theta(x)$ in \eqref{eq:1}. However, the problem is that we need to compute the evidence, $P_\theta(x)$ in the below Bayes' theorem, to calculate the posterior in each step.  

\subsubsection{Variational Autoencoder}

Rather than sampling from $P_\theta(x)$ directly, we can approximate the posterior with the recognition model $q_\phi(z|x)$, and use the neural network to optimize $q_\phi(z|x)$. Then, this structure becomes similar to autoencoder. \cite{kingma2013auto} propose the VAE which optimizes the variance lower bound using the neural network and uses the reparameterization trick to backpropagate the loss.

In VAE, we update the $q_\phi(z|x)$ instead of the posterior. Therefore, we have to minimize the difference between $q_\phi(z|x)$ and $p(z|x)$, while maximizing the likelihood of $p(x)$. Therefore, the objective function of VAE is to maximize the log-likelihood of $p_\theta(x)$ while minimizing the KL-divergence between $q_\phi(z|x)$ and $p_\theta(z|x)$, and this is variational lower bound of $\theta$ and $\phi$.

\begin{align}
    L(\theta, \phi) &= \log p_\theta(x) - D_{KL}(q_\phi(z|x)\,||\,p_\theta(z|x))\nonumber \\
    &= \int dz \;q_\phi(z|x) \log \frac{p_\theta(x,z)}{q_\phi(z|x)}\nonumber \\
    &= \mathop{\mathbb{E}}_{z\sim q_\phi(z|x)}[\log p_\theta(x|z)] - D_{KL}(q_\phi(z|x)||p_\theta(z))\label{eq:2}
\end{align}

Therefore, optimizing the variance lower bound is the same as to minimize the sum of the reconstruction error - $\mathop{\mathbb{E}}_{z\sim q_\phi(z|x)}[\log p_\theta(x|z)]$ and the regularized error - $D_{KL}(q(z|x)||p(z))$. Using this relation, we can increase the variance lower bound in every iteration by optimizing $\theta, \phi$ by backpropagation. However, for backpropagation to work properly, in \eqref{eq:2}, we still need to get the derivative for the sampling.

\begin{figure}[ht]
    \centering
    \includegraphics[width=6cm]{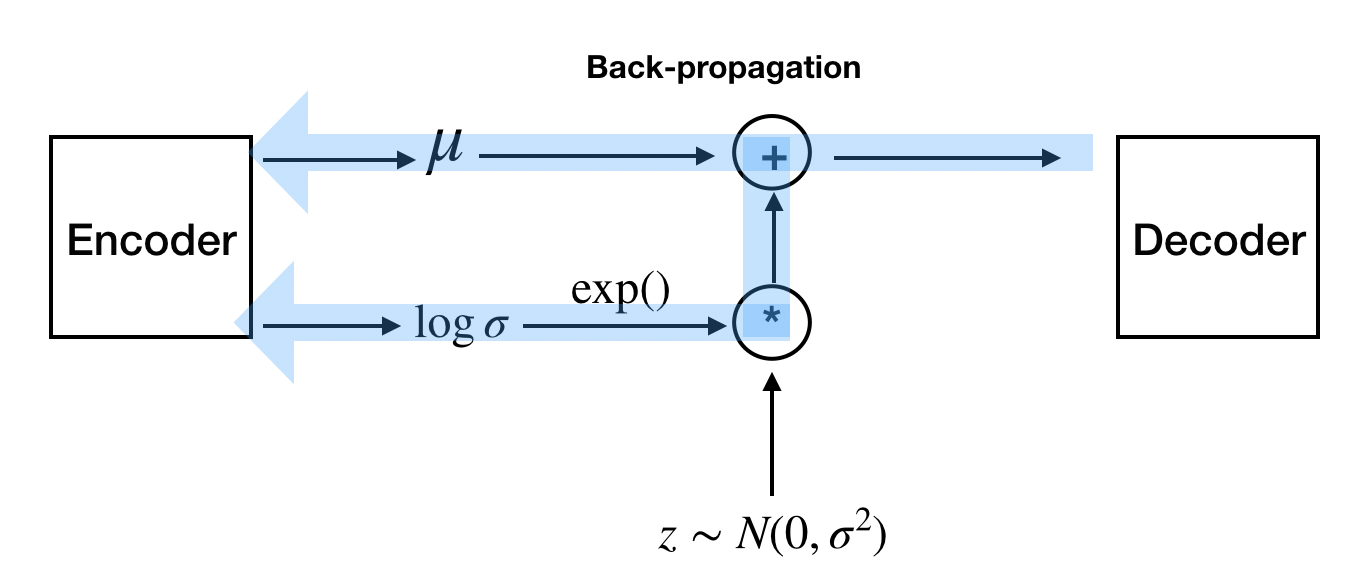}
    \caption{Reparameterization Trick in VAE}
    \label{fig:1}
\end{figure}

The authors cleverly circumvent the problem of getting the gradient of sampling by using the reparameterization trick. The idea is to get the mean vector and the covariance matrix from Encoder, and they are the mean and the covariance matrix of $q_\phi(z|x)$. Then, we get the sample from standard Normal distribution, and multiply the covariance matrix and then add the mean vector. By using the reparameterization trick, we can get the gradients of mean and covariance without having to worry about the sampling. To implement the VAE in real application, we usually take the $\log\;\sigma$ and take the exponential value of it. Then, the output from the Encoder is not necessarily restricted to the positive values. It is illustrated in the Figure \ref{fig:1}.

Since we assume that the encoded latent distribution($q_\phi(z|x)$) follows Normal distribution, we can compute $D_{KL}(q_\phi(z|x)||p(z))$ in the closed form. Then, we can compute the regularized error exactly as below. For the reconstruction error which is $\mathop{\mathbb{E}}_{z\sim q_\phi(z|x)}[\log p_\theta(x|z)]$, we need to minimize the negative log-likelihood or any kind of the loss function defined over the output.

\begin{align}
    D_{KL}\;(q(z|x)||p(z))\;=\;-\frac{1}{2}\sum_j(1+\log((\sigma_j)^2)-(\mu_j)^2-(\sigma_j)^2)\nonumber
\end{align}

VAE has the solid theoretical background, and training VAE tends to be relatively stable. Also, if we use proper conditional variables along with the input data, we can explain the source of variations easily like semi-supervised learning method. However, there were some criticisms. The first criticism is that why we sample from the standard Normal distribution which has diagonal covariance matrix. It is more natural to assume that the covariance matrix is not diagonal.

Another criticism usually comes from the element-wise loss functions. It leads the lower quality of generated images. The lower quality of images comes from the fact that element-wise loss does not capture the spatial correlation between the features. Usually, for reconstruction loss, the \textit{L2} loss function is used, but it averages the different data distributions, so the image becomes a bit blurred.

\subsubsection{Flow models and Disentanglement}

There have been approaches to use conditional distributions and normalized flows for posterior distributions. Then, we can improve the VAE model by approximating the posterior distribution having non-diagonal covariance matrices. Also, if we assume that the $q_phi(z)$ follows the autoregressive relation, then it is scalable because the covariance matrix will be triangular matrices as in \eqref{eq:3}.(\cite{kingma2016improved}) 

\begin{align}
    \log q(z_T|x) = \log q(z_0|x) - \displaystyle\sum_{t=1}^T \log \det |\frac{dz_t}{dz_{t-1}}| \nonumber
\end{align}

Also, there have been other attempts to change the regularized error on VAE. \cite{makhzani2015adversarial} proposed adversarial autoencoder which uses Discriminator to classify the encoded latent distribution and the pure Gaussian. \cite{higgins2017beta} proposed $\beta$-VAE to penalty the KL divergence away from the specific value $\beta$. However, larger $\beta$ is more likely to make the latent space factorized, but it also reduces the quality of the outputs. \cite{kim2018disentangling} improves $\beta$-VAE model by using the additional adversarial classifier which determines the latent space is properly factorized or not instead of restricting the regularized error by $\beta$. 

\subsubsection{VAE + GAN}

After VAE and GAN have been proposed, many papers are trying to hybrid these two models. The basic idea is to use the VAE structure for more stable training and to obtain the adversarial loss to improve the quality of outputs. 

\cite{ulyanov2017adversarial} proposed Adversarial Generative-Encoder(AGE) networks which incorporated adversarial loss inside the system. By corporating the adversarial loss, it minimizes the adversarial loss on the latent space and the reconstruction loss on the output space in VAE. The benefit of this structure is that it does not require another critic network to classify which one is generated.

\cite{rosca2017variational} proposed $\alpha$-GAN which uses two Discriminators to decide whether the latent encoding is from the original or not and whether the image is the output or generated. The authors claim that Density Ratio well approximates the KL divergence between the posterior $q(x|z)$ and the prior $p(z)$, and use Density Ratio trick to describe the loss function.

\cite{huang2018introvae} proposed IntroVAE which modifies the loss functions of AGE. The authors change the loss function to incorporate the \textit{minmax} game structure between the generated output from the standard Gaussian and the original image at the Encoder, and the generated output and the input at the Decoder.

\subsection{Generative Adversarial Network}

Along with VAE, recently there has been the new generative modeling approach called GAN developed by \cite{goodfellow2014generative}. It is inspired by the game theory. There are two networks in GAN, one is Generator, and the other is Discriminator, and they play the game. Generator generates the samples from the noise and tries to fool the Discriminator that the generated samples are real. Meanwhile, Discriminator seeks to increase the probability of classifying the real example. Therefore, the two networks play \textit{minimax} game, and the Nash Equilibrium of these two networks would be the probability of 0.5 for discriminating the real and the generated images. For Generator(G) and Discriminator(D), the loss function would be below.

\begin{align}
    \min_G\;\max_D L(D,G)\;\;=\; \mathbb{E}_{x \sim p_r(x)}\;[\log D(x)]\;+\;\mathbb{E}_{z \sim p_z(z)}\;[\,\log\;(1-D(G(z)))]\nonumber
\end{align}

If the GAN is properly trained and reaches the global optimal point, the loss function would be $-2\log 2$, and $p_r$ = $p_g$, and $D(x) = \frac{1}{2}$. At the optimal point, the loss function would be the Jenson-Shannon(JS) divergence between the probability distribution of samples and the data distribution of the real data.

GAN is known to produce very sharp images if it is trained well, but the training of GAN is notoriously difficult and unstable. First of all, it is tough to reach the Nash Equilibrium just updating parameters through gradient descent algorithm. The Nash Equilibrium can be achieved only through the \textit{cooperative behaviour of these two networks}, but in plain GAN, two models update themselves independently. This does not guarantee the convergence of these two networks.

Secondly, it is not easy for the networks to have proper gradients to train. If the discriminator performs well, then their gradient would be quite small (because it is close to one), so the generator does not have the proper feedback. In another case, if the discriminator doesn't behave properly, the gradient would be quite large, but this feedback is not appropriate to train the generator. This is well explained in Figure 2.

\begin{figure}[ht]
    \centering
    \includegraphics[width=6cm,height=3cm]{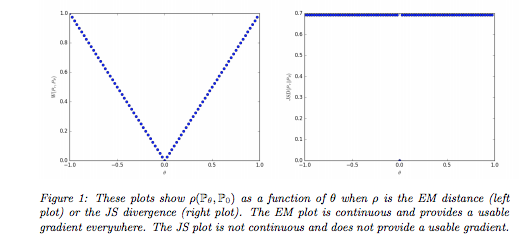}\label{fig:2}
    \caption{Image from \cite{arjovsky2017wasserstein}}
\end{figure}

Moreover, according to \textit{manifold learning hypothesis}, data usually lie in small latent dimensions even if data has artificially large dimensions. Therefore, the probability distribution of real data and the generated data would be located on tiny dimensions. In other words, it means that they are highly likely to be disjoint. If they are disjoint, then KL divergence between the two distribution is infinity, and the JS divergence is the constant, so it does not have the proper gradient to update the parameters. On top of this, during the training, GAN usually settles in suboptimal equilibrium and keeps producing the same outputs to fool the discriminator. This settling of generating similar images is called the Mode Collapse.

\subsubsection{Improved GAN training, WGAN, WGAN-GP}

To resolve the problems mentioned above, many approaches have been introduced in the last few years. \cite{salimans2016improved} proposed some techniques for training GANs. They include feature mapping, virtual batch normalization, minibatch discrimination, label smoothing and history averaging, and adding noises. 

\cite{arjovsky2017wasserstein} proposed the new metric for GAN. Rather than using JS divergence, the authors suggested using Wasserstein-1 Distance (WD). WD, also known as Earth Mover's distance, is a measure of distances between two distributions. GAN using this WD metric is called Wasserstein GAN (WGAN). It tries to find the most efficient way of transporting one distribution to another distribution $p_g$.

\begin{align}
    W(p_r,p_g)\;=\; \inf_{\gamma\sim\Pi(p_r,p_g)} \mathop\mathbb{E}_{(x,y)\sim\gamma}[||x-y||]\label{eq:3}
\end{align}

WD is useful metric because the distance between two distributions is just the difference between the two distributions. Unlike KL or JS divergences, the gradient based on WD depends on how far away the two distributions are. Therefore, the role of the Discriminator changes in WGAN because the Discriminator measures how far the generated and the real distributions are, so the authors rename Discriminator as Critic. Also, it avoids the mode-collapse problem which networks stuck in sub-optimal equilibrium by cooperative behavior between Discriminator and Critic However, finding the optimal transport in \eqref{eq:3} is intractable, so by Kantorovich-Rubinstein Duality, the equation could be rewritten as \eqref{eq:4}.

\begin{align}
    W(p_r,p_g) = \sup_{||f||_L\leq K}\mathbb{E}_{x \sim P_r}[f(x)]-\mathbb{E}_{x \sim p_g}[f(x)]\label{eq:4}
\end{align}

Kantorovich-Rubinstein Duality is only achieved when the function meets K-Lipschitz condition(in \eqref{eq:4},$||f||_L\leq K$, K is hyperparameter here). Therefore, the distributions need to meet this condition. \cite{arjovsky2017wasserstein} proposed to clip the parameters which the authors admit that it is not a good approach, and \cite{gulrajani2017improved} came up with giving the penalty to the gradient if the gradient measured on the sample point between two distributions is larger than 1. This is WGAN-GP, and we can rewrite the equation as \eqref{eq:5}

\begin{align}
    L=\mathbb{E}_{x\sim p_g}\;[D(x)]-\mathbb{E}_{x\sim p_r}\;[D(x)] + \lambda\mathbb{E}_{x'\sim p_g}[(\|\nabla_{x'}\;D(x')\|^2-1)^2]\label{eq:5}
\end{align}

The $\lambda$ in \eqref{eq:5} is the hyperparameter, and it is usually set to 10. Even though this penalized term increases the complexity of the model, but WGAN-GP enlarges the capacity of WGAN.

\subsection{Colorization}

Colorization has been a challenging question in the area of computer vision and machine learning. It is challenging because the gray-scale image corresponds to several colorful images, so it means that there are more than one correct answer. There have been several approaches to this task after the neural network is used widely in the field of computer vision.

\cite{cheng2015deep} approached the colorization task as the regression problem. They used CNN with grey-scale input and YUV-color space output. The authors use least-square minimization as the loss function.

\begin{figure}[ht]
    \centering
    \includegraphics[width=8cm]{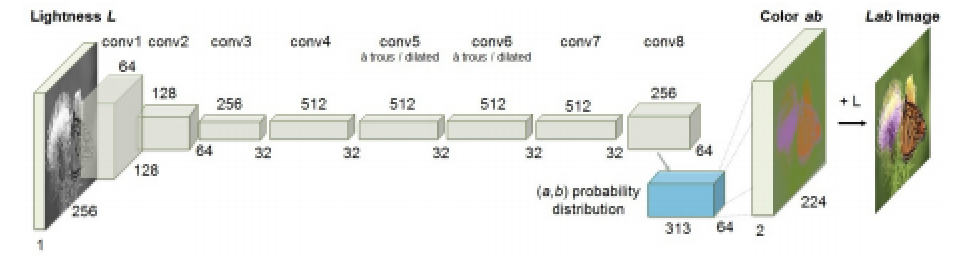}
    \caption{From \cite{zhang2016colorful}}
\end{figure}

\cite{zhang2016colorful} proposed CNN structure to colorize the grey-scale image. The authors used the \textit{Lab} space rather than RGB, and they classified Lab space into 313 \textit{ab} pairs, and apply multinomial cross entropy loss function to represent the inherent multi-modality in colorization. Also, the authors rebalance the training set data considering the different populations of colors in the training data. They trained on ImageNet dataset, and for measuring the performance of the output, the authors proposed the Human-Turing test.

\cite{deshpande2017learning} proposed VAE structure along with Mixture Density Network(MDN). MDN can helps the function correspond to several different outputs structurally. They feed the grey-scale image as conditional input to MDN, and original images into VAE. They modify the L2 loss function by the weighted sum of the Mahalanobis distance of Top-20 Principal components in the image. They used the data set of Labelled Faces in the Wild dataset, LSUN-Church, and ImageNet-Validation.

\begin{figure}[ht]
    \centering
    \includegraphics[width=8cm,height=4cm]{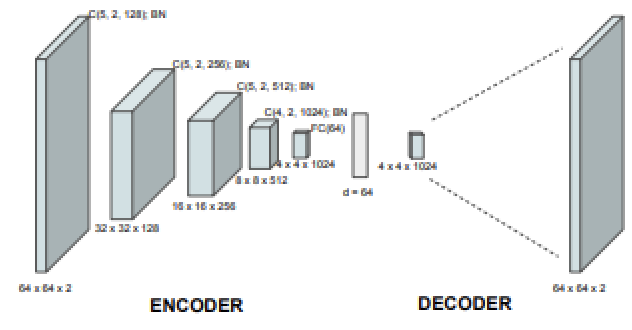}
    \caption{From \cite{deshpande2017learning}}
\end{figure}

\cite{isola2017image} used conditional GAN structure to colorize the image. They applied the U-Net architecture as shown in the below figure. Conditional GAN and conditional VAE both use the information bottleneck structure and it is known to lose some information about the detail. However, U-Net successfully circumvents the problem of information bottleneck. It also incorporates L1 loss function which is known to preserve the spatial structure, so it is a bit different from a vanilla GAN and more like a hybrid structure between VAE and GAN. 

\begin{figure}[ht]
    \centering
    \includegraphics[width=8cm]{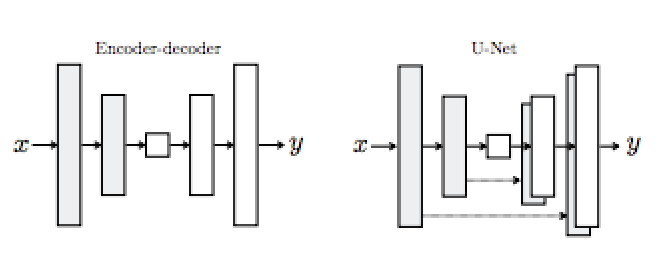}
    \caption{From \cite{isola2017image}}
\end{figure}

There have been attempts to colorize the grey-scale images for all different neural network architectures. Previous methods try to tackle the multi-modality of colorization by changing the loss function or using generative modelings. The problem is how to penalize the plausible different color of the ground truth images. Also, previous papers usually depend on a human's eye to compare the results with the ground truth.  

\subsection{Evaluation Metrics}

\cite{salimans2016improved} proposed the Inception Score (IS) to measure the performance of generative modeling. The basic idea of IS is that IS should be highly confident if there is a single clear object in the image, and the outputs are highly diversified. To incorporate the aforementioned idea, $p(y|X)$ should be low entropy and $p(y)=\int_x\;p(y|x)p(x)$ should be high entropy. To measure the entropy, the authors use the pre-trained model to calculate its outputs at the specific hidden layer. The authors claim that it correlates well with human eyes, and it is specifically well fitted to CIFAR-10 images.

\begin{align}
    IS(G) = \exp(E_{x\sim p_g}D_{KL}(p(y|x)\|p(y)))\nonumber
\end{align}

As we have seen the above equation, IS is KL divergence between $p(y|x)$ and $p(y)$. Therefore, if $p(y|x)$ is sharp, and $p(y)$ is wide enough to cover all the regions, IS would be high. To compute the IS, we use the pre-trained Inception network and get the values from the intermediate layers. These values are considered as the features from the images, and use those images to compute IS.

IS is a widely used metric, but it also has some drawbacks. Many different outputs generate the same IS, so it can be fooled if the gradient is tuned well. Also, it cannot classify the original image from the generated images. \cite{heusel2017gans} proposed Frechet Inception Distance(FID) which captures the similarity of generated images. FID uses the pre-trained network to extract features from intermediate layers, then model the features using the multivariate Gaussian distributions. FID is quite sensitive to mode collapse problems. Unlike IS, FID is better if the values are less.

\begin{align}
    FID(x,g)=\|\mu_x-\mu_g\|^2_2+\text{Trace}(\Sigma_x+\Sigma_g-2(\Sigma_x\Sigma_g)^{\frac{1}{2}})\nonumber
\end{align}

\subsection{The setting of the Experiments}

I have trained each model up to 100 epoch using CIFAR-10 training data set. Then, I compute the IS using the generated images and generate some of the images using the grey-scale CIFAR-10 images. I have converted CIFAR-10 images using the \textit{scikit-image rgb2gray} function. Then, I generated colorized images through different models. Original CIFAR-10 images achieve the IS of 9.3. 

\section{Convolutional Neural Network}

The perfornamce of CNN is the baseline of this colorization task. The basic idea of this model is to feed the grey-scale image as the input, and generate the colorful images as output, and compute the loss function by comparing the generated colorful images with the original image. Its structure is modified from \cite{zhang2016colorful}, but \cite{zhang2016colorful} uses \textit{Lab} scale images and uses multinomial loss function in \textit{a,b} space. Here, I use RGB space, and L1 \& L2 reconstruction loss functions.

\subsection{The structure of the model}

In this structure, the gray-scale input has less information than the colorful images. Rather than using the information bottleneck style structure, I maintained the total number of dimensions while I downsampled using convolutions with strides 2. I also used dilated Convolution to increase the receptive field large enough without having to use pooling or linear layers.

\begin{table}[ht]
\begin{center}
\begin{tabular}{|c|c|c|c|c|c|c|c|c|}
\hline
Layer & 1 & 2 & 3 & 4 & 5 & 6 & 7 & 8 \\ \hline\hline
Convolution & 3x3 & 3x3 & 3x3 & 3x3 & 3x3 & \begin{tabular}[c]{@{}c@{}}3x3(T)\end{tabular} & \begin{tabular}[c]{@{}c@{}}3x3(T)\end{tabular} & \begin{tabular}[c]{@{}c@{}}3x3(T)\end{tabular} \\ \hline
Channels Out & 16 & 64 & 256 & 256 & 256 & 128 & 64 & 3 \\ \hline
Stride & 2 & 2 & 2 & 1 & 1 & 2 & 2 & 2 \\ \hline
Padding & 1 & 1 & 1 & 2 & 2 & 1 & 1 & 1 \\ \hline
Dilation & 1 & 1 & 1 & 2 & 2 & 1 & 1 & 1 \\ \hline
Output padding & - & - & - & - & - & 1 & 1 & 1 \\ \hline
Activation & \begin{tabular}[c]{@{}c@{}}Leaky\\ ReLU\end{tabular} & \begin{tabular}[c]{@{}c@{}}Leaky\\ ReLU\end{tabular} & \begin{tabular}[c]{@{}c@{}}Leaky\\ ReLU\end{tabular} & \begin{tabular}[c]{@{}c@{}}Leaky\\ ReLU\end{tabular} & \begin{tabular}[c]{@{}c@{}}Leaky\\ ReLU\end{tabular} & \begin{tabular}[c]{@{}c@{}}Leaky\\ ReLU\end{tabular} & \begin{tabular}[c]{@{}c@{}}Leaky\\ ReLU\end{tabular} & Tanh \\ \hline
Batch Norm & Yes & Yes & Yes & Yes & Yes & Yes & Yes & No \\ \hline
\end{tabular}
\caption{The overall structure of CNN(T for Transposed Convolution)}
\end{center}
\label{table:1}
\end{table}

I choose kernel size as three because there would be five convolution layers in sequence with the first three convolution layers having strides 2 and the remaining two having stride 1 and dilation of 2. Therefore, the five layers work as encoding layers, and they have enough receptive fields (74) to recognize the input images. The last three transposed convolution layers work as decoding layers.

For the padding, I calculate the number of padding dimensions to maintain the dimension of the images same after the convolution. By doing the padding \textit{same}, the network can become more in-depth.

For each convolution layer, the activation function is LeakyReLU(0.2) is followed. I found out that ReLU activation sometimes suffers gradient vanishing problems when there are negative inputs, so decided to use LeakyReLU and with the parameter 0.2 for the negative inputs without adding any parameters. The last activation function is \textit{Tanh} function because the original data has the range of -1 to 1.  

For loss function, I use mean-square error and do not include any regularized terms. Also, I train the model using L1 loss and compare the result with the model using L2 loss. L1 loss function is widely used in the field of computer vision because it is known to produce less blurry images on edges. For optimizer, Adam Optimizer with the fixed learning rate \textit{0.001} worked better with compared to Stochastic Gradient Descent with the same learning rate.

\subsection{Performance of CNN}

As we have seen the above, CNN directly transforms the grey-scale images into the colorized images. It takes two hours to train the model having the architecture of table \ref{table:1} with CIFAR-10 images respectively for both L1 and L2 loss up to 100 epochs using \textit{Amazon AWS p3.xlarge}. To compute the IS, I use test images of CIFAR-10.

\begin{figure}[ht]
    \centering
    \includegraphics[width=10cm]{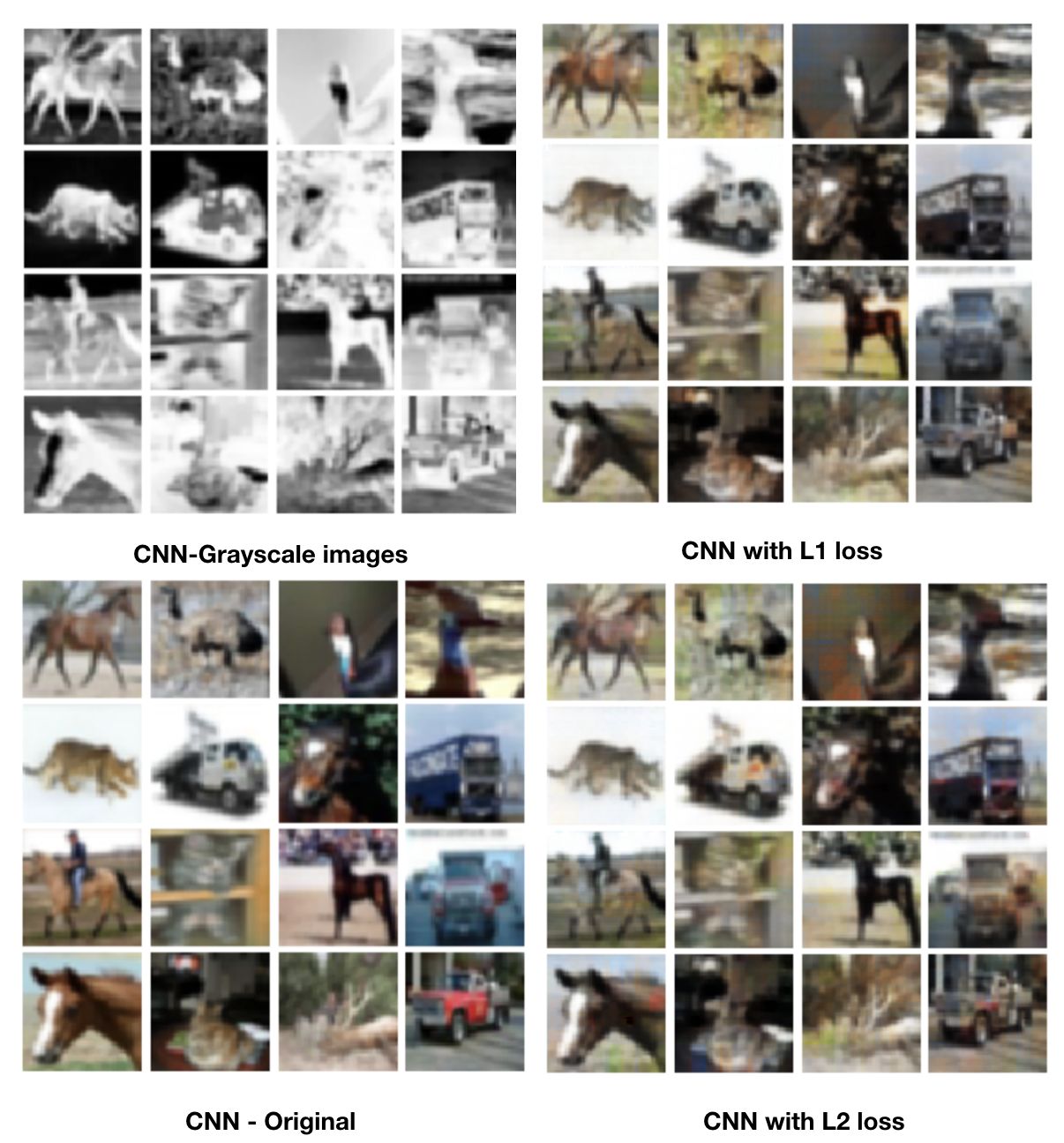}\label{fig:6}
    \caption{Selected samples of generated pictures and original images}
\end{figure}

As you see the above Figure \ref{fig:6}, original images are a little bit more natural but it is not easy to recognize the big difference just by looking at it. 

\begin{figure}[ht]
    \centering
    \includegraphics[width=8cm]{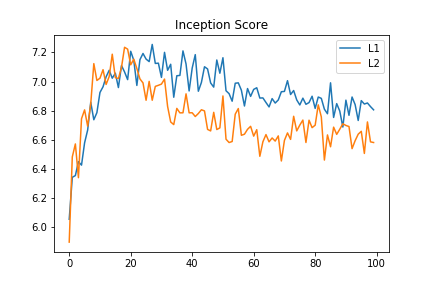}
    \caption{The Inception Score of CNN with L1 and L2 losses for each epoch}
\end{figure}

The IS of two models reach the peak around 20-30 epochs, and the overall trend is likely to decrease after the peak. It stems from the fact that both models tend to overfit since they do not have the regularized loss. Also, as we have seen from the above graph, CNN with L1 loss tends to have higher IS. This high Inception Score is plausible because L1 loss tends to generate less blurry images, and IS increases as the images are more distinctive.

\section{Conditional VAE}

For this colorization task, I use the grey-scale image as a conditional input to VAE along with colorized images. Therefore, after the training, Decoder generates colorized images only by receiving the conditional grey-scale images.

\subsection{The structure of the model}

Variational lower bound also holds for CVAE just by adding the conditional term on the equation. For training the neural network, using the proper auxilliary variable greatly stabilizes the process not only for VAE but also for GAN.

\begin{align}
    L(\theta, \phi|c) &= \mathop{\mathbb{E}}_{z\sim q_\phi(z|x,c)}[\log p_\theta(x|z,c)] - D_{KL}(q_\phi(z|x,c)||p_\theta(z))\label{eq:6}
\end{align}

The purpose of encoder at CVAE is to understand the image. Therefore, I use 4 series of convolution layers followed by LeakyReLU activation and Batch Normalization, except the last layer and the first convolution layer has 5x5 filters in order to have the large receptive fields. 

\begin{table}[ht]
\begin{center}
\begin{tabular}{|c|c|c|c|c|c|}
\hline
Layer & 1 & 2 & 3 & 4-1 & 4-2 \\ \hline\hline
Convolution & 5x5 & 3x3 & 3x3 & (Dense) & (Dense) \\ \hline
Channel Out & 16 & 64 & 256 & 512 & 512 \\ \hline
Stride & 2 & 2 & 2 &  &  \\ \hline
Padding & 1 & 1 & 1 &  &  \\ \hline
Dilation & 1 & 1 & 1 &  &  \\ \hline
Activation & \begin{tabular}[c]{@{}c@{}}Leaky\\ ReLU\end{tabular} & \begin{tabular}[c]{@{}c@{}}Leaky\\ ReLU\end{tabular} & \begin{tabular}[c]{@{}c@{}}Leaky\\ ReLU\end{tabular} & Linear & Linear \\ \hline
Batch Norm & Yes & Yes & Yes & No & No  \\ \hline
\end{tabular}
\caption{The Structure of Encoder(CVAE)}
\end{center}
\end{table}

For conditional part, I use the three set of convolution layers also followed by LeakyReLU and Batch Normalization for gray-scale images except the last convolution layer, while \cite{deshpande2017learning} uses MDN for gray-scale image to encourage the VAE to generate diverse images. Here, the first convolution layer has larger fiter than other layers in order to have enough receptive fields. Also, when I downsample with convolution layer with strides 2, I increase the channel at least by 4 (the first time by 16) to make sure that Conditional part understands the full semantics of the gray-scale images.

\begin{table}[ht]
\begin{center}
\begin{tabular}{|c|c|c|c|}
\hline
Layer & 1 & 2 & 3 \\ \hline\hline
Convolution & 5x5 & 3x3 & 3x3 \\ \hline
Channel Out & 16 & 64 & 256 \\ \hline
Stride & 2 & 2 & 2 \\ \hline
Padding & 1 & 1 & 1 \\ \hline
Dilation & 1 & 1 & 1 \\ \hline
Activation & \begin{tabular}[c]{@{}c@{}}Leaky\\ ReLU\end{tabular} & \begin{tabular}[c]{@{}c@{}}Leaky\\ ReLU\end{tabular} & \begin{tabular}[c]{@{}c@{}}Leaky\\ ReLU\end{tabular} \\ \hline
Batch Norm & Yes & Yes & No \\ \hline
\end{tabular}
\caption{The Structure of Conditional(CVAE)}
\end{center}
\end{table}

Decoder receives three inputs - two($\mu$ and log $\sigma$) from Encoder and one tensor featured from Conditional which has the grey-scale input. First, the reparameterisation trick is applied to both $\mu$ and log $\sigma$. Sample \textit{z} from standard normal distribution, and add $\mu$ and multiply exponential log $\sigma$. The latent dimension of this Encoder is 512, and it has enough model capacity to generate images similar to ground-truth. Also, WGAN-GP and WGAN-GP with L1 loss also have 512 dimensions.

\begin{table}[ht]
\begin{tabular}{|c|c|c|c|c|c|c|c|c|c|}
\hline
Layer & 1 & 2 & 3 & 4 & 5 & 6 & 7 & 8 & 9 \\ \hline\hline
Kernel & Re & Dense & \begin{tabular}[c]{@{}c@{}}Reshape/\\ Concat\end{tabular} & \begin{tabular}[c]{@{}c@{}}3x3\\ Conv\end{tabular} & \begin{tabular}[c]{@{}c@{}}3x3\\ Conv\end{tabular} & \begin{tabular}[c]{@{}c@{}}3x3\\ Conv\end{tabular} & \begin{tabular}[c]{@{}c@{}}3x3(T)\\ Conv\end{tabular} & \begin{tabular}[c]{@{}c@{}}3x3(T)\\ Conv\end{tabular} & \begin{tabular}[c]{@{}c@{}}3x3(T)\\ Conv\end{tabular} \\ \hline
\begin{tabular}[c]{@{}c@{}}Output\\ Dimension\end{tabular} & 512 & 4096 & (8,8,128) & 256 & 256 & 256 & 128 & 64 & 3 \\ \hline
Stride & - & - & - & 1 & 2 & 2 & 2 & 2 & 2 \\ \hline
Padding & - & - & - & 1 & 2 & 2 & 1 & 1 & 1 \\ \hline
Dilation & - & - & - & 1 & 2 & 2 & 1 & 1 & 1 \\ \hline
\begin{tabular}[c]{@{}c@{}}Output\\ Padding\end{tabular} & - & - & - & - & - & - & 1 & 1 & 1 \\ \hline
Activation & - & \begin{tabular}[c]{@{}c@{}}Leaky\\ ReLU\end{tabular} & - & \begin{tabular}[c]{@{}c@{}}Leaky\\ ReLU\end{tabular} & \begin{tabular}[c]{@{}c@{}}Leaky\\ ReLU\end{tabular} & \begin{tabular}[c]{@{}c@{}}Leaky\\ ReLU\end{tabular} & \begin{tabular}[c]{@{}c@{}}Leaky\\ ReLU\end{tabular} & \begin{tabular}[c]{@{}c@{}}Leaky\\ ReLU\end{tabular} & Tanh \\ \hline
\begin{tabular}[c]{@{}c@{}}Batch\\ Normalisation\end{tabular} & - & No & - & Yes & Yes & Yes & Yes & Yes & No \\ \hline
\end{tabular}
\caption{The structure of Decoder(CVAE), Re: Reparameterisation Trick, T:Transposed}
\end{table}

After the reparameterisation trick, the transformed sample is fed into the dense layer with the output size of 4096, and it is reshaped into an 8x8x64 tensor. Then, the output of the dense layer which is from the standard Gaussian would be concatenated to the output of the Conditional layer which also has 8x8x64 dimension, so now the concatenated tensor has 8x8x128 dimension. This would be fed into all convolutional, dilated convolutional and transposed convolutional blocks which are followed by LeakyReLU activation and Batch Normalisation except the last layer which has Tanh activation and no Batch Normalisation. I intentionally make the latter part of the Decoder similar to CNN for comparison.

For reconstruction loss, I use L1 loss and mean-square error respectively without averaging the size. For regularized loss, I use the KL-divergence of the posterior and the standard Gaussian prior derived by \cite{kingma2013auto}. Through the training process, the neural network tries to minimize the combined error of reconstruction and regularized error. I also use Adam Optimizer with the fixed learning rate of 0.001 and betas = (0.5, 0.999).

\subsection{Performance of CVAE}

In CVAE, it generates the colorful images given the ground-truth images and the grey-scale conditional images. The IS of CVAE with L1 and L2 losses are higher than those of CNN with both L1 and L2 loss. It takes five hours to train each model up to 100 epochs with \textit{Amazon AWS p3.2xlarge}.

The generated images are very close to ground-truth images except for some small details. To compare CVAE with L1 and CVAE with the L2 loss, CVAE with L1 can generate subtle differences in color for the different semantics. VAE is known to generate blurry images, but the generated images here look very distinctive. The intermediate results under one epoch and 50 epochs are shown below.

\begin{figure}[ht]
    \centering
    \includegraphics[width=10cm]{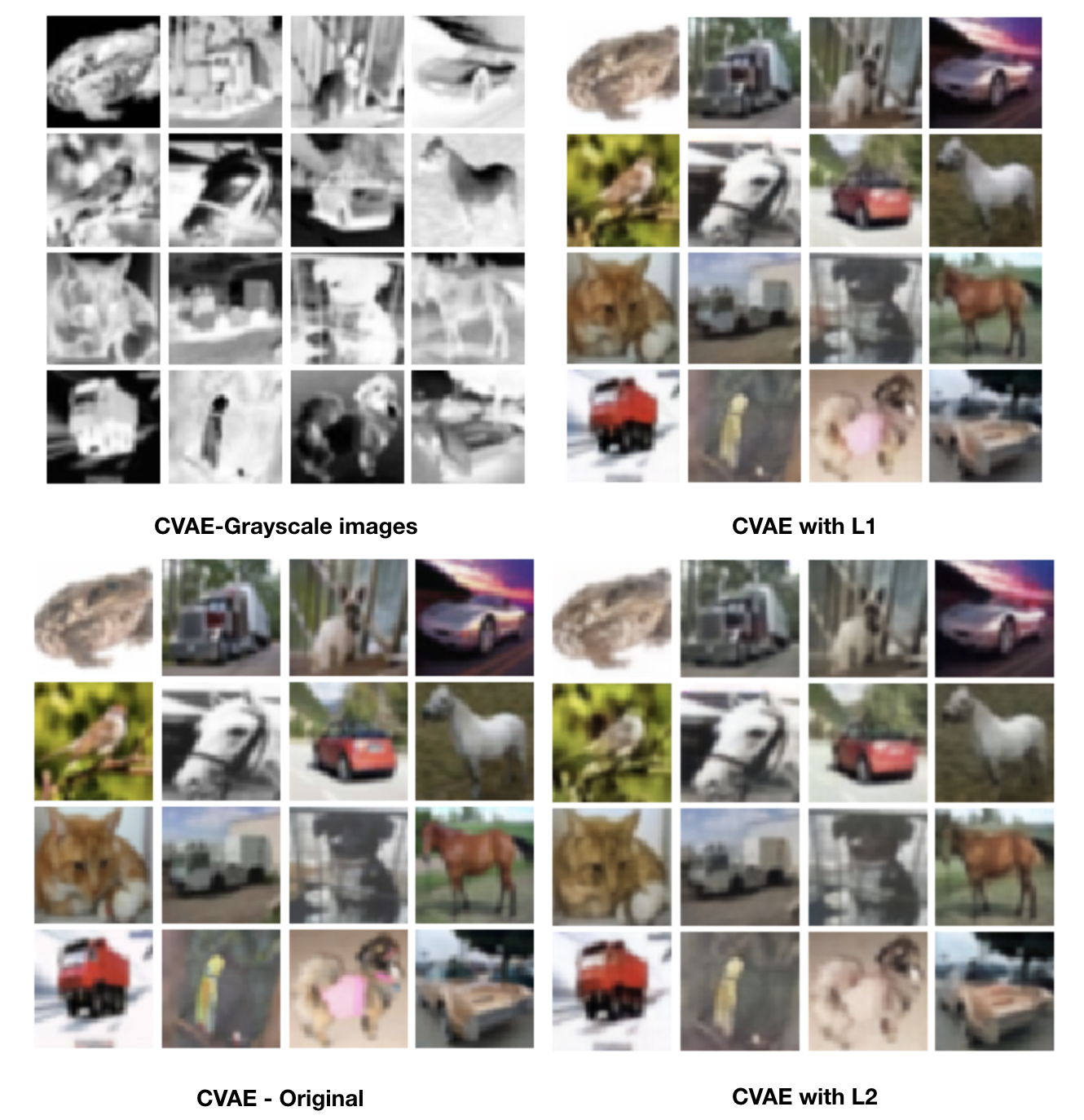}\label{fig:7}
    \caption{Selected samples of generated pictures and original images}
\end{figure}

The IS of CVAE with both L1 and L2 loss achieve very high Inception Score. CVAE with L1 reconstruction loss has slightly higher Inception Score than that with the L2 loss, also consistent with the results from CNN. When we compare this with the CNN, we can see that IS does not decrease even after both reach plateau after 50 epochs. Unlike CNN, CVAE does not suffer from over-fitting. This means that regularized error which is KL divergence between the posterior $q(z|x)$ and $p(z)$ is quite useful to prevent overfitting. Also, the values of IS are a bit larger than that of IS achieved by \cite{gulrajani2017improved} where the most of the methods reach the IS between 6 and 7.

\begin{figure}[ht]
    \centering
    \includegraphics[width=8cm]{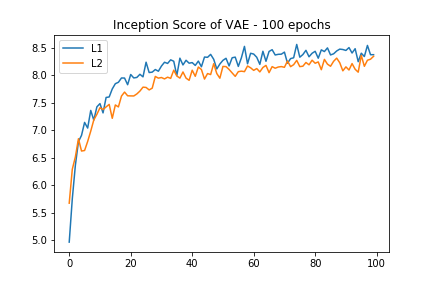}
    \caption{The Inception Score of CVAE with L1 and L2 losses for each epoch}
\end{figure}

\section{CWGAN-GP and CWGAN-GP with L1 reconstruction loss}

To colorize the gray-scale images with GAN structure, Conditional GAN would be the reasonable choice like CVAE. CWGAN-GP has better capacity and is more stable to train than CWGAN. Also, I will compare CWGAN-GP with CWGAN-GP with L1 reconstruction loss and explore how L1 reconstruction error and WD change as the network is trained. Both CWGAN-GP and CWGAN-GP with L1 loss have some architecture and only the loss function is different.

\subsection{The Structure of the models}

I feed the grey-scale image as input to Generator. Also, I add the grey-scale image as a conditional variable to Critic to improve the training. In overall, as we can see from the above figure, the grey-scale image is fed into Generator as conditional input.

CWGAN-GP is also not easy to train the model, so I have attempted many tips as suggested in both papers and blogs. One of the most effective way is to feed the gray-scale image also into Critic. Critic becomes more stable and learns better with this additional input. It may be because Critic understands the semantics of the input better with its gray-scale image, and provide better feedback to Generator to be properly trained.

\subsubsection{Conditional WGAN-GP}

For Generator, it receives the grey-scale images as conditional input and generates the colorized images as output. For Generator, I made the structure similar to CVAE as I mentioned the above, but the different thing is that I combine the \textit{Conditional} and \textit{Decoder} inside the \textit{Generator}.

\begin{figure}[ht]
    \centering
    \includegraphics[width=10cm]{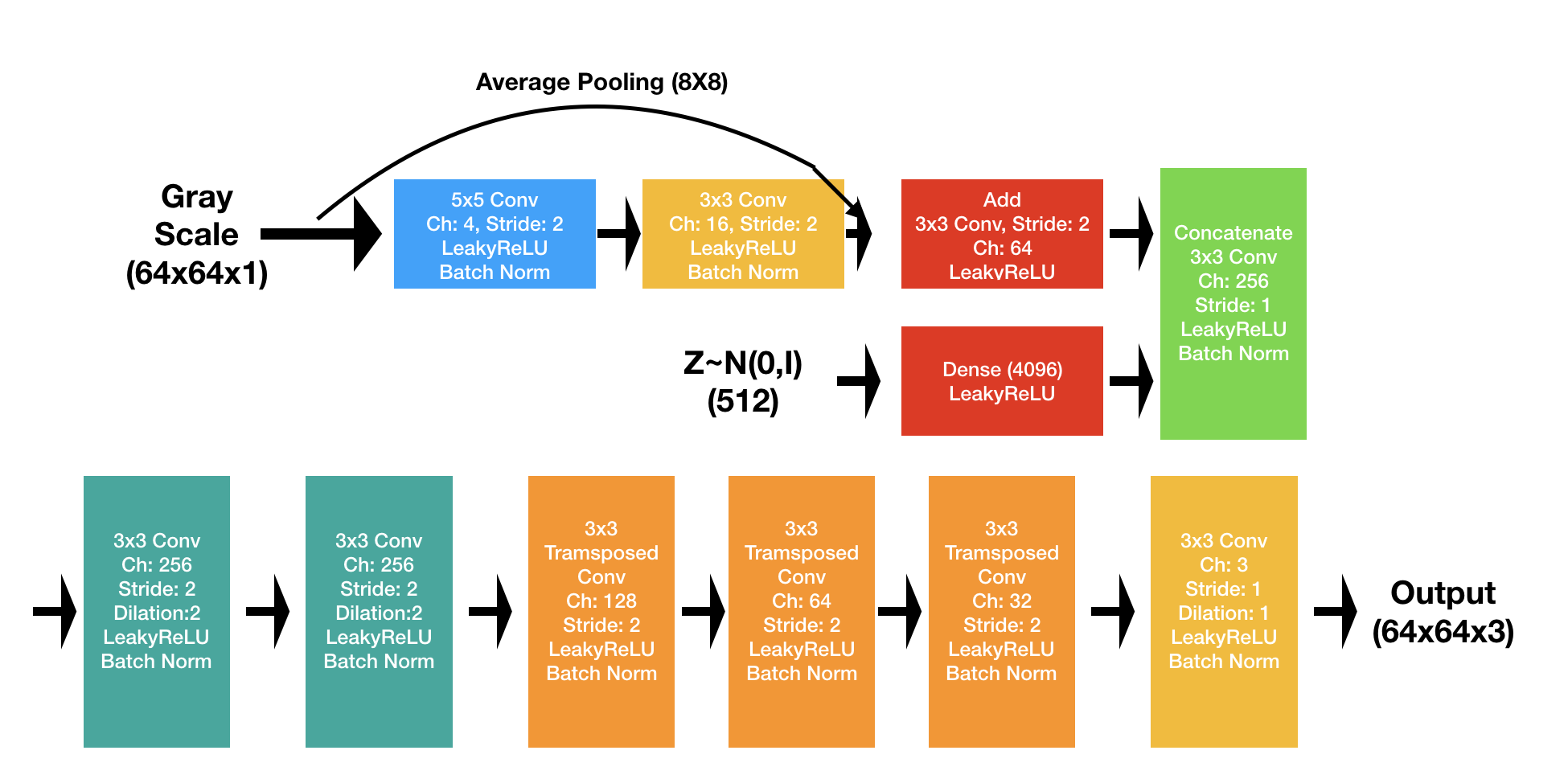}
    \caption{Overall structure of Generator}
\end{figure}

I also use residual connection when I extract features from grey-scale images. By making a residual connection, Generator extracts the feature from the first two convolution layers better. It improves the performance of the Generator in overall. 

However, after concatenating the sample from the standard Gaussian and the features from grey-scale images, I do not apply residual connection. I found out that the residual connection is good at keeping gradients and data from previous layers, but it may affect adversely for generating new information in the following layers. 
For Critic, as I mentioned the above, it receives two inputs - one grey-scale image and one ground-truth or generated image. Then, I apply two residual blocks and one average pooling and the other dense layer to get 1-dimension vector for each image. This structure is motivated by \cite{gulrajani2017improved} which uses residual connections in Critics for their implementation on the CIFAR-10 data set.

\begin{figure}[ht]
    \centering
    \includegraphics[width=10cm]{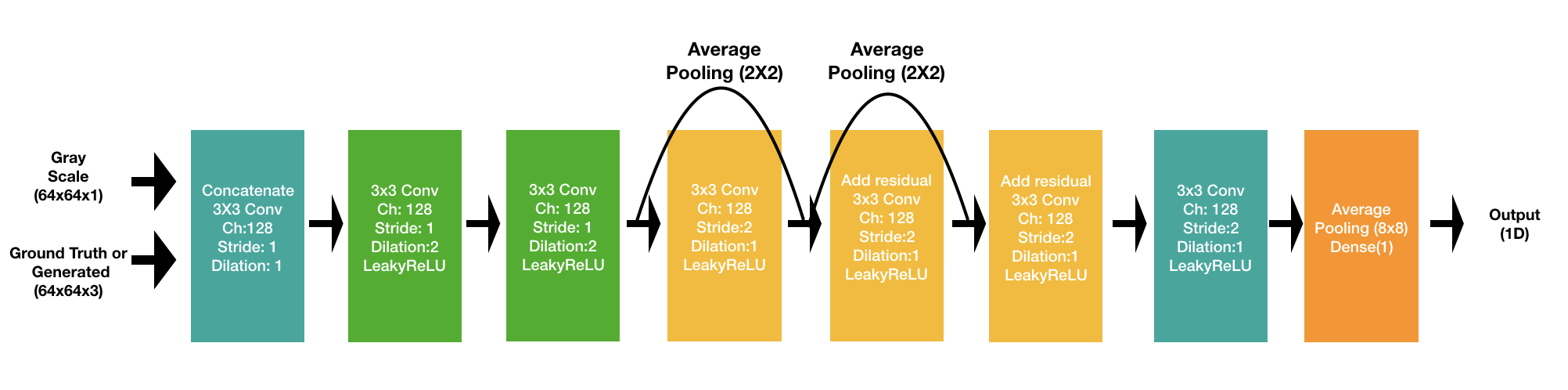}
    \caption{Overall structure of Critic}
\end{figure}

I use three convolution layers with stride 1, which means that there is no downsampling, to have enough receptive fields to understand the semantics of the images. Then, I use two residual connections in the middle to keep the features extracted in the first three layers during downsampling. Then, I use one convolution layer with the kernel size 3 to downsample once more, then do pooling on average and use the dense layer to get 1-dimension vector. I use the linear activation layer in the end to get the WD between two distributions.

One of the crucial things in designing the Critic is not to use Batch Normalization. Batch Normalization tends to correlate between samples, and it adversely affects the performance of the Critic. Therefore, I do not use any normalization techniques for training. Instead, I use residual connections in the middle to keep the information extracted from the first three convolution layers with stride 1, and do not use max-pooling but average-pooling to get the overall characteristics of the images.

To implement Gradient Penalty, I sample the point between the real image and the generated image, and get the output from the Critic with the sampled point, and get the gradient of this point using the \textit{autograd} feature of Pytorch. Then, get the difference of the value 1 and the gradient of this value, and obtain the square of the value as mentioned above, then multiply ten which is hyper-parameter. \cite{gulrajani2017improved} recommended to use $\lambda$ as 10 and it works well for this process too.

From my experiments, Gradient Penalty reduces the WD of two distributions. I also implemented WGAN with clipping the weights, but in this case, the performance of WGAN varies a lot with the value of clipping, and the training became unstable.

There are two kinds of losses in WGAN-GP. Critic loss is the negative value of WD(the difference the means of the output of Critics between the generated image and the ground truth) plus the Gradient Penalty of the sample between the real image and the generated image which means the gradient of the example should be close to 1 by assuming 1-Lipschitz condition. In other words, Critic loss is \textit{the penalty} which violates the 1-Lipschitz condition of the sample between Real and Generated images \textit{plus the WD} distance of the real and the generated images. Therefore, if the distribution of generated images does not violate the 1-Lipschitz condition and it is close to the distribution of real images, then the Critic loss would be minimized.

Generator loss is the negative value of the mean of the output of Critics of the generated image, and this value would be considered as of how unlikely the Critic classifies the generated images. Therefore, the neural networks try to minimize the Generator loss in order not to be classified as fake.

\subsubsection{Conditional WGAN-GP with L1 reconstruction loss}

This method is actually motivated by \cite{isola2017image}. In CWGAN-GP, the loss functions are computed over the outputs of Critic. However, we may improve this model by also adding the L1 difference between the generated output and the input at the Generator stage. The total loss would be the summation of the loss of CWGAN-GP and the L1 reconstruction loss.

The same structure of Generator and Critic in CWGAN-GP is applied for CWGAN-GP with L1 reconstruction loss, and the loss function is updated by adding the L1 distance between the Ground Truth and the generated images. This would help the Generator train from ground-truth images.

However, L1 reconstruction loss and the adversarial loss from the Critic are different. It would be interesting to see the relationship between WD and L1 loss during the training. The training ratio of Critic and Generator is 1:1 in this case because the Generator can also get the feedback from L1 loss, so it can be trained well even if the Critic is not yet trained enough. Also, same Adam Optimizer with linearly interpolated learning ratio from 0.001 to 0 over 100 epochs is used.

\subsection{Performance of CWGAN-GP and CWGAN-GP with L1 loss}

In CWGAN-GP, the grey-scale images are fed into both Generator and Critic, and the colorized images are generated from Generator, and Critic receives the generated samples or real images and measures how far the distributions of both data and samples. It takes 26 hours to train CWGAN-GP up to 100 epochs with \textit{Amazon AWS p3.2xlarge}. CWGAN-GP with L1 reconstruction loss has the same structure as CWGAN-GP, and the only difference is that it also minimizes the L1 difference between the generated output and the real input, and the training ratio between the Generator and the Critic is 1 to 1. It takes 30 hours to train CWGAN-GP with L1 loss up to 100 epochs with \textit{Amazon AWS p3.xlarge}.

\begin{figure}[ht]
    \centering
    \includegraphics[width=8cm]{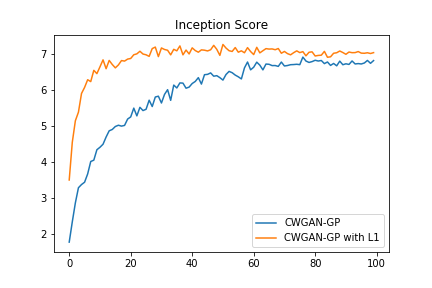}
    \caption{The Inception Score of Conditional WGAN-GP}
\end{figure}

The IS of CWGAN-GP and CWGAN-GP with L1 loss reach around 7. Both models do not increase beyond Inception Score of 7.5. It probably comes from the fact that the structure of GAN does not have \textit{Encoder}, so it has limited capacity to generate the input images entirely. The learning curve of CWGAN-GP with L1 loss is steeper than CWGAN-GP in the earlier stage. This comes from the fact that Generator in CWGAN-GP with L1 loss is trained to minimize the L1 loss together. However, it seems that this L1 loss does not increase the performance of the model significantly.

\begin{figure}[ht]
    \centering
    \includegraphics[width=10cm]{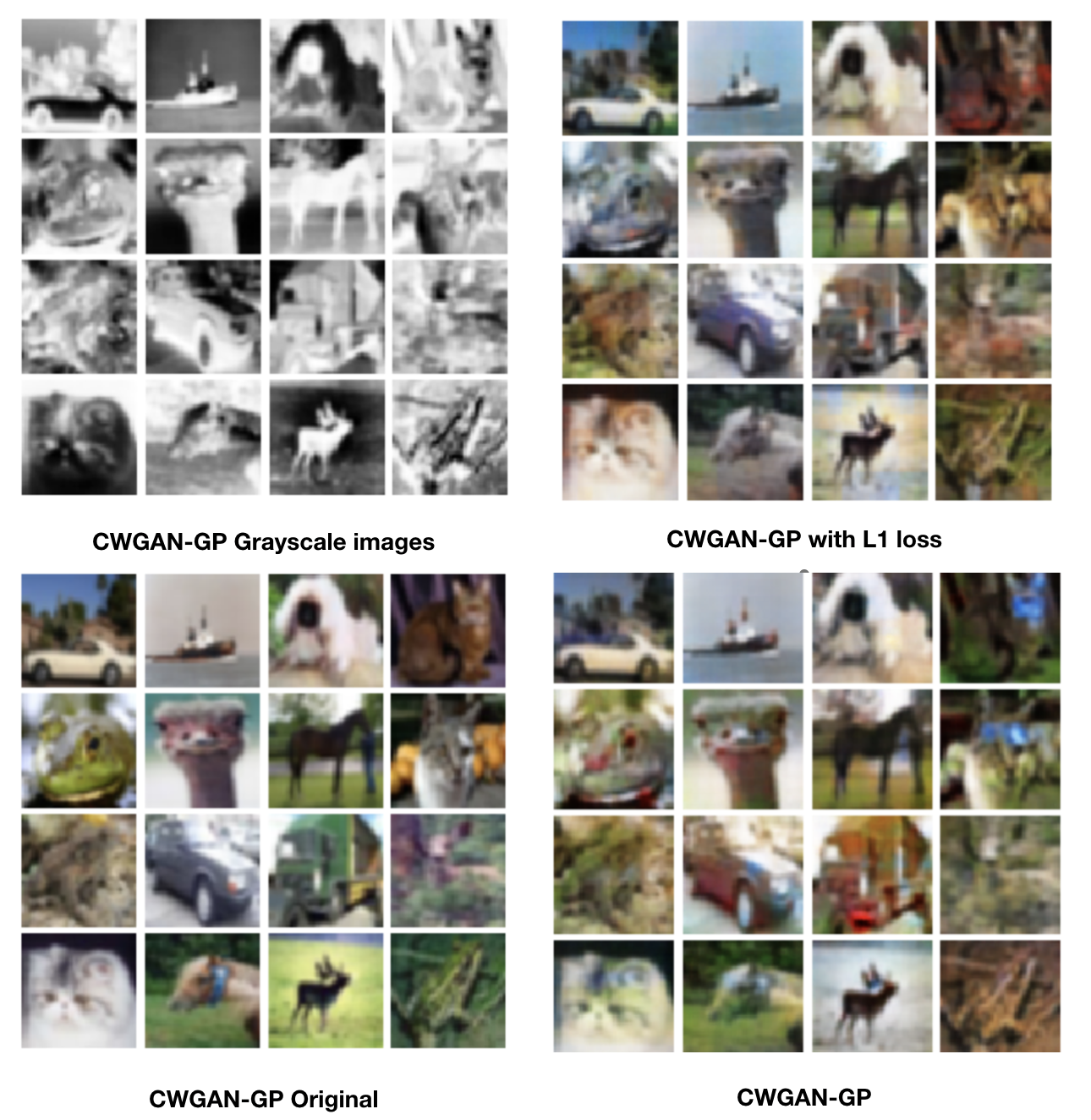}\label{fig:8}
    \caption{Selected samples of generated pictures and original images(CWGAN-GP)}
\end{figure}

As we have seen the pictures in Figure \ref{fig:8}, CWGAN-GP still has some weird color mappings as you can see some bluish faces on the upper right rabbit pictures. CWGAN-GP tends to generate more diverse images, and this comes from the adversarial loss structure.

\begin{figure}[ht]
    \centering
    \includegraphics[width=10cm]{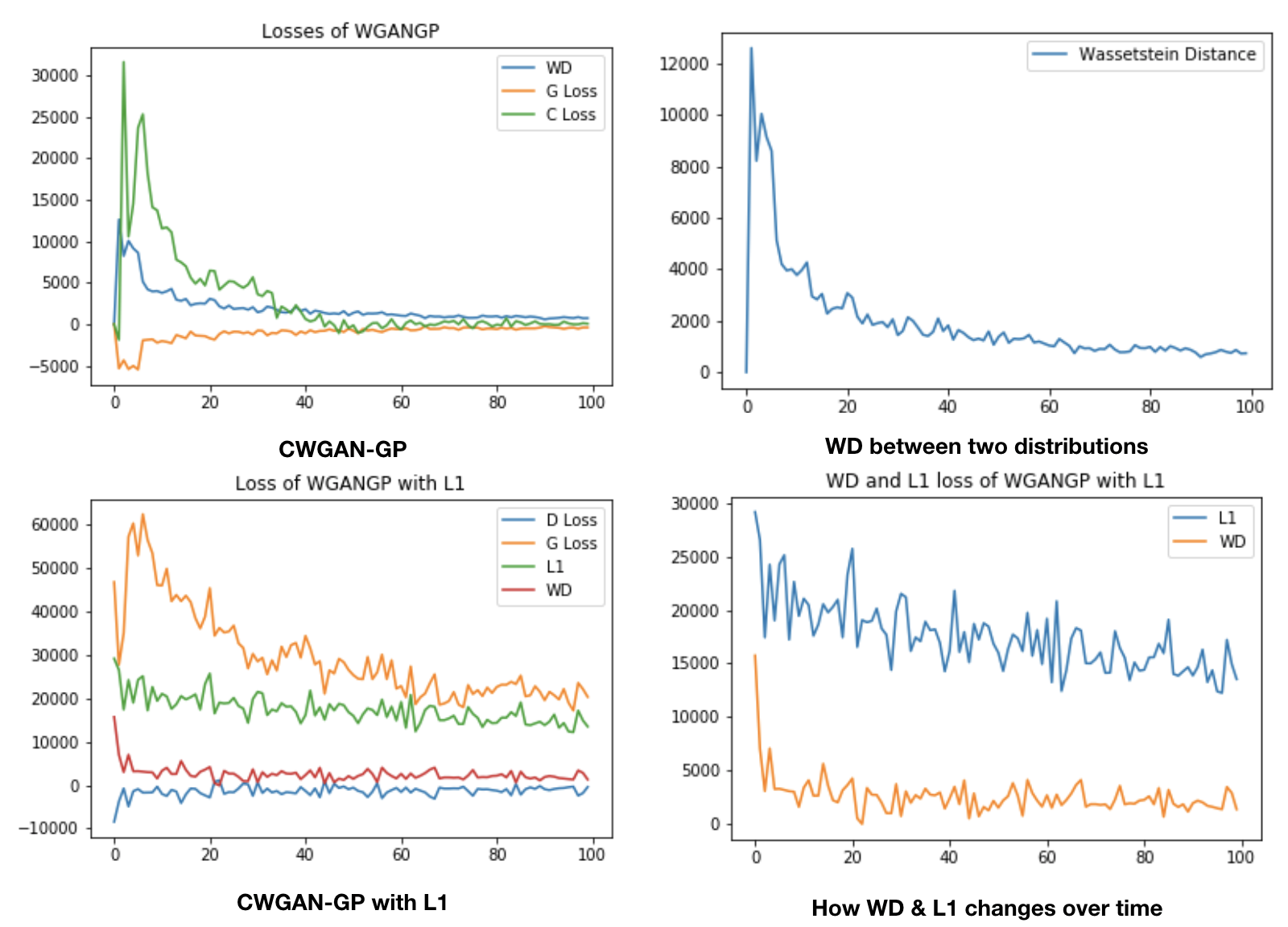}\label{fig:9}
    \caption{Loss for CWGAN-GP and CWGAN-GP with L1 loss}
\end{figure}

As we can see from Figure \ref{fig:9}, the WD increases in the early few epochs, and then it keeps going down until 100 epochs. In Figure 5.6, the L1 loss tends to go down in the entire training period, while WD becomes stable after ten epochs. These patterns clearly show that CWGAN-GP with L1 reconstruction loss tends to learn faster but get stuck around Inception Score of 7, whereas CWGAN-GP tends to learn the features through the entire training epochs, and they reach the similar results in the end.

\section{VAE+GAN models}

In this section, I am going to present some of the VAE+GAN models - AGE by \cite{ulyanov2017adversarial} and IVAE by \cite{huang2018introvae}. WGAN-GP with L1 reconstruction loss also can be considered as one of VAE+GAN models which also minimizes the reconstruction loss as well as adversarial loss. As we've seen before, this does not necessarily increase the model capacity though it helps the network to learn faster. AGE and IVAE are different because they change the loss function similar to \textit{minmax} game between the two networks. I have modified AGE and IVAE for colorization task. Here, I use L1 loss because it produces the better results as we have seen before. Also, both use the same architecture of CVAE and change the loss function and how to update parameters.

\subsection{Adversarial Generative Encoder}

The overall structure of AGE is to sample the standard Normal distribution, then generate the image through the Decoder, and then encode the generated image and the real image together then get the L1 distance between the two, and then feed the generated encoding to Decoder and get the generated image again and minimize the L1 distance between the second generated image and the real image.

In other words, AGE minimizes the distance in \textit{latent space} between the generated image and the real image, so that Encoder cannot differentiate between the real image and the generated image from the standard Normal. Then, AGE also minimizes the distance in \textit{output space} between the generated output and the real output.

\begin{algorithm}[ht]
\caption{Adversarial Generative Encoder}
$Z\; \sim\; N(0,1)\\
X_{gen}\; \leftarrow \;Dec(Z)\\
Z_{gen}\; \leftarrow \;Enc(X_{gen})\\
Z_{real}\; \leftarrow \;Enc(X_{real}) \\
L_{adv}\; \leftarrow\;\text{L1}\;(Z_{gen}, Z_{real})\\
\theta \leftarrow \theta - \alpha \nabla_{\theta} (L_{adv})\\
X_{gen}\; \leftarrow \; Dec(Z_{gen})\\
L_{rec} \; \leftarrow \; \text{L1}\;(X_{gen}, X_{real})\\
\phi \leftarrow \phi - \beta\nabla_{\phi} (L_{rec})$
\end{algorithm}

For loss function, I use L1 loss both in the latent dimension and in the output dimension. Adam Optimizer with learning rate 0.001 and betas = (0.5, 0.999) are used for Encoder, Conditional and Decoder.

\subsection{Introspective VAE}

The idea of IVAE is also to incorporate the \textit{minmax} game structure by changing the loss function like AGE. It generates the sample through Decoder twice - one from standard Gaussian and the other from the latent encodings of the generated images of standard Gaussian.  

\begin{algorithm}[ht]
\caption{Introspective VAE(\textit{ng} means no gradient)}
$Z \leftarrow Enc(X) \\
Z_p\; \sim\; N(0,1)\\
X_r \leftarrow Dec(Z) \\
X_p \leftarrow Dec(Z_p) \\
L_{AE} \leftarrow L_{AE} (X_r, X)\\
Z_r \leftarrow Enc(ng(X_r)), Z_{pp} \leftarrow Enc(ng(X_p))\\
L^E_{adv} \leftarrow L_{REG}(Z) + \alpha{[m-L_{REG}]^{+} + [m-L_{REG}]^{+}}\\
\phi_E \leftarrow \phi_E-\eta \nabla_{\phi_E}(L_{adv}^E + \beta L_{AE})\\
Z_r  \leftarrow Enc(X_r), Z_{pp} \leftarrow Enc(X_p)\\
L^G_{adv} \leftarrow \alpha {L_{REG}(Z_r) + L_{REG}(Z_{pp})}\\
\theta_G \leftarrow \theta_G - \eta \nabla_{theta_G}(L^G_{adv} + \beta L_{AE})$
\end{algorithm}

Two types of loss functions are defined in IVAE, one is an adversarial loss on the encoder, and the other is an adversarial loss on the decoder. IVAE samples \textit{the first output} from Decoder directly using the standard Gaussian samples and gets the encoded latent information of the above samples. Then, IVAE also gets \textit{the second output} again by using the encoded latent information from the standard Gaussians. The adversarial loss on the encoder is minimized when the latent encodings from ground-truth are close to both \textit{the first output} and \textit{the second output}. The adversarial loss on the decoder is to minimize the encodings of \textit{the first output} and \textit{the second output}. 

I set the hyper-parameters in the above Algorithm as $\alpha$ = 0.5 and $\beta$ = 0.4 (in the above Algorithm). I set $\alpha$=0.5 and m = 120 because adversarial losses are less than the threshold(m). If they are less than the threshold, $\alpha$ = 0.5 represents the difference between the latent encoding of the original and the generated images. If $\beta$ becomes too small, the reconstruction error becomes small in the total loss. Therefore, I set $\beta$ = 0.4 because it provides the highest IS.

IVAE also uses the same structure as CVAE, and change the loss functions as described the above. The Adam Optimizer with the fixed learning rate of 0.001 and betas=(0.5, 0.999) are used for Encoder, Conditional and Decoder. For IVAE, I only measure L1 loss because L1 is known to produce the better results.

\subsection{Performance of AGE and IVAE}

In AGE and IVAE, both have the same structure as CVAE in chapter 4, and they have only different loss function. Both models take approximately seven hours to train AGE up to 100 epochs with \textit{Amazon p3.2xlarge}.

\begin{figure}[ht]
    \centering
    \includegraphics[width=8cm]{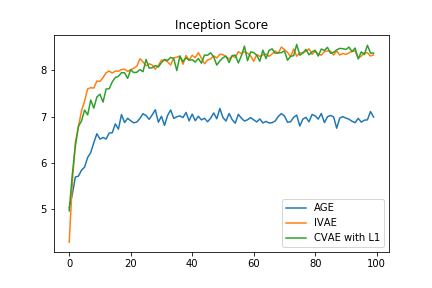}
    \caption{The IS of AGE and IVAE with compared to CVAE with L1 loss}
\end{figure}

The IS of AGE saturates around 7, but that of IVAE reaches about 8.4, and it is quite compatible with that of CVAE with the L1 loss. IVAE could achieve the almost similar result with compared to CVAE.

\begin{figure}[ht]
    \centering
    \includegraphics[width=8cm]{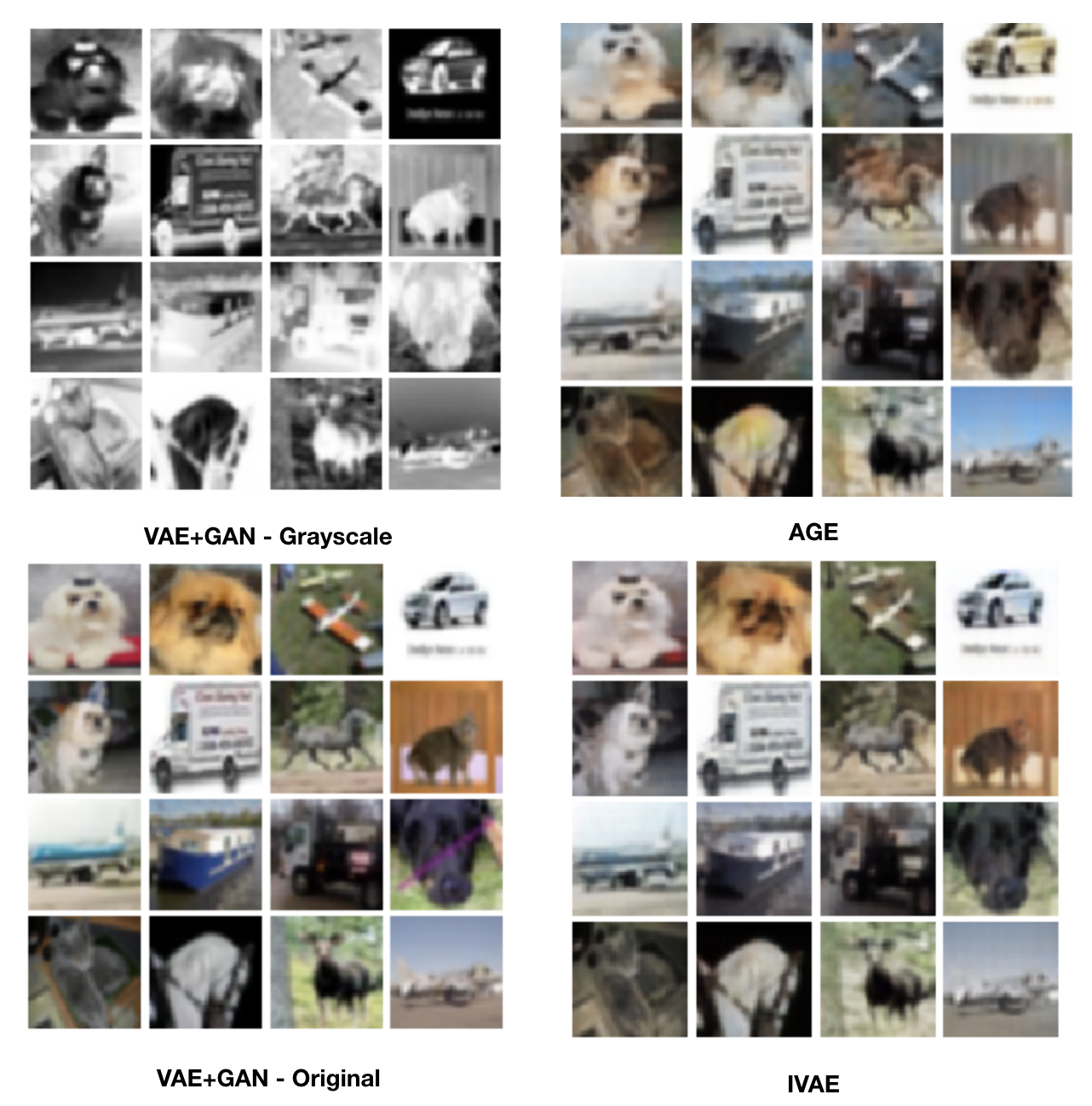}
    \caption{Selected samples from AGE and IVAE}
\end{figure}

\section{Conclusions}

In this paper, we explore some generative modelings for colorization task. By comparing the value of IS, CVAE and IVAE with L1 loss produce the best results. The L1 loss tends to produce better results than the L2 loss in both CNN and CVAE. CWGAN-GP with L1 loss learns a bit faster than CWGAN-GP, but both models reach IS around 7. Both AGE and IVAE incorporate the adversarial loss inside the VAE structure by changing the loss function, but their IS is entirely different, and CNN tends to overfit without regularizer. 

One interesting to notice is that CWGAN-GP with L1 loss and AGE achieve the similar IS of 7, but the L1 distance of AGE is less than that of CWGAN-GP with L1 reconstruction loss. The adversarial loss within CWGAN-GP makes this difference between AGE and CWGAN-GP with L1 construction loss

\begin{figure}[ht]
    \centering
    \includegraphics[width=8cm]{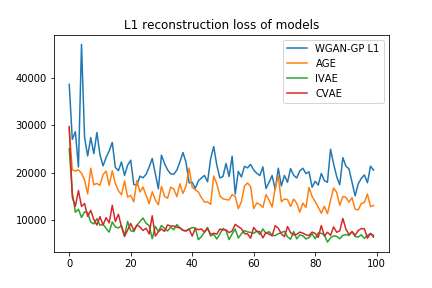}
    \caption{L1 loss of models using L1 loss}
\end{figure}

For training time, it takes more than one day to train for CWGAN-GP and CWGAN-GP with L1 reconstruction loss. This is primarily because both use more complicated structures. 

\begin{table}[ht]
\begin{tabular}{|c|c|c|c|c|c|c|}
\hline
\begin{tabular}[c]{@{}c@{}}Models\\ (Amazon p3.2xlarge)\end{tabular} & CNN & CVAE & CWGAN-GP & \begin{tabular}[c]{@{}c@{}}CWGAN-GP\\ with L1\end{tabular} & AGE & IVAE \\ \hline
\begin{tabular}[c]{@{}c@{}}Hours\\ (For 100 epochs)\end{tabular} & 2 & 5 & 26 & 30 & 7 & 7 \\ \hline
\end{tabular}
\end{table}

To sum up, the several important points during training are as below. For CNN, dilated convolution layers tend to increase not only the receptive field of the kernels but also the quality of the outputs. For CVAE, conditioning on proper variables are quite useful to stabilize the training, and the KL divergence between the posterior and the standard Gaussian is very useful as well as theoretically sound. For CWGAN-GP, checker-point dots are less likely to be observed if the model is adequately trained, and also the learning curve should become smaller as two networks get trained more. Using the conditional variable on Discriminator is very useful for the entire networks to learn better. For CWGAN-GP with the L1 loss, I realized that minimizing the distance on the output space does not necessarily lead to minimize the WD between two distributions. For IVAE and AGE, changing the loss function similar to \textit{minimax} game only may not be enough to incorporate VAE and GAN. For future works, I would like to develop CWGAN-GP more for training more stable.

\bibliographystyle{plainnat}
\bibliography{references}
\end{document}